\documentclass{article}
\usepackage{epsfig}
\usepackage{amssymb}

\newcommand{\etal}{{\em et~al.}}              

\newcommand{\plb}[1]{Phys.~Lett.~B {#1}}

\newcommand{\npa}[1]{Nucl.~Phys.~{A#1}}
\newcommand{\prc}[1]{Phys.~Rev.~C {#1}}

\newcommand{\beq}{\begin{equation}}
\newcommand{\eeq}{\end{equation}}
\newcommand{\bea}{\begin{eqnarray}}
\newcommand{\eea}{\end{eqnarray}}

\begin{document}
\title{Bulk Nuclear Properties from Reactions}
\author{P.\ Danielewicz \\
\rm National Superconducting Cyclotron Laboratory and\\
 Department of Physics and Astronomy,\\
Michigan State University, East Lansing, MI 48824, USA
}

\maketitle

\begin{abstract}
Extraction of bulk nuclear properties by
comparing reaction
observables to results from semiclassical
transport-model simulations is discussed.
Specific properties include the nuclear viscosity,
incompressibility and constraints on the nuclear pressure
at supranormal densities.
\end{abstract}

\section{Introduction}
I shall discuss the extraction of bulk nuclear properties from
reactions.  Of particular interest will be central reactions of
heavy nuclei, characterized by a multitude of
emitted particles and by a multitude of competing physical
effects.  These reactions are commonly described in terms of
phase-space distributions $f$ that follow the Boltzmann
equation:
\bea
\nonumber
{\partial f \over \partial t} + {\partial \epsilon_{\bf p} \over
\partial {\bf p} }
\, {\partial f \over \partial {\bf r}} -
{\partial \epsilon_{\bf p} \over
\partial {\bf r} } \, {\partial f \over \partial {\bf p}}
& = &
\int { d{\bf p}_2 }
\int d \Omega' \,
v_{12} \, \frac{d \sigma}{d \Omega'} \,
 \big( (1 - f_1) (1 - f_2)  \\
&& \times  f_1' \, f_2'  - (1 - f_1') (1 - f_2') f_1 \, f_2 \big) \, .
\label{Boltzmann}
\eea
Here, $\epsilon({\bf p}, \lbrace f \rbrace) $ is the single
particle energy.  The~terms on the l.h.s.\ of the equation
account for the changes of $f$ due to the motion of particles
in the average potential field produced by other particles;
the particle velocity is ${\bf v} = \partial \epsilon /
\partial {\bf p}$.  The
r.h.s.\ of (\ref{Boltzmann}) accounts for changes of $f$ due to
collisions.

The transport relying on (\ref{Boltzmann}) has been quite successful in
applications, describing a multitude of
measured single-particle spectra, among other.  With a confidence stemming
from the success of predictions,
the transport theory allows for
a good insight into the history and mechanism of
reactions.  The theory is fairly flexible allowing
one to include new particles as energy domain changes and to
incorporate new collision processes if these become
important.

Despite successes of the theory, there are significant
uncertainties in the underlying Boltzmann equation.  Thus, the
dependence of the single-particle energies on momentum and
density is generally not known.  In terms of the net system
energy, the single-particle energies are:
\beq
 \epsilon = \frac{\delta E}{ \delta f} \, ,
\eeq
and they relate to particle optical potentials with
\beq
U_{opt} = \epsilon - \epsilon_{kin} \, .
\eeq
The cross sections utilized in
the collision integral in (1) are usually such as in
free-space,
but different cross sections may need to be utilized in the
medium.  An issue may be the very validity of the Boltzmann
equation in a dense system.  What if the theory is only
phenomenological?

The indicated uncertainities represent difficulties but also
opportunities to learn about nuclear systems.
The opportunities related to uncertainties include e.g. the
nuclear
equation of state (EOS) generally related to single-particle
energies,
and the nuclear transport coefficients related to
in-medium cross sections.
For progress, it
is necessary
to identify observables from reactions, or combinations
thereof, that are sensitive to a specific uncertainty.
It is
necessary to understand which particular features of the
nuclear system are explored in a reaction and why an outcome
may be well described in spite of the uncertainties.
In the
following, I shall give examples of the inference of bulk
properties of nuclear matter from comparing the transport
results to reaction data, emphasizing the above points.

\section{Stopping, Cross Sections and Viscosity}

Observables describing stopping of nuclei on each other in
reactions might be used to extract information on in-medium
cross sections.  However, are the cross sections an objective
characterization of a system?  Can one talk about isolated
collisions when the medium is dense and excited so that Pauli
principle does not suppress the frequency of collisions?
What about macroscopic system properties?  Within the Boltzmann
description, the cross-sections are related to the viscosity,
proportional to the mean free path and inversely proportional
to the cross sections:
\bea
\nonumber
\eta & = & {5 \over 9} \, T \, \left[ \int d{\bf p} \, p^2 \, f
\right]^2  \left/ \int d{\bf p}_1 \int d{\bf p}_2 \int
d\Omega'
\,
v_{12} {d \sigma
\over
d \Omega'} \, q_{12}^4 \sin^2 \, \theta'
\right.
\\ && \times
f_1 \, f_2 \, (1-f_1') \, (1 - f_2')
\, ,
\label{eta}
\eea
with the relative momentum equal to $q_{12}= | {\bf p}_1 - {\bf
p}_2|/2$.  When manipulating the cross sections within
simulations,
one alters the nuclear viscosity and one can hope that stopping
observables probe the viscosity even while a link to cross
sections remains ambiguous.

\begin{figure}
\centerline{\includegraphics[angle=0,
width=.67\linewidth]{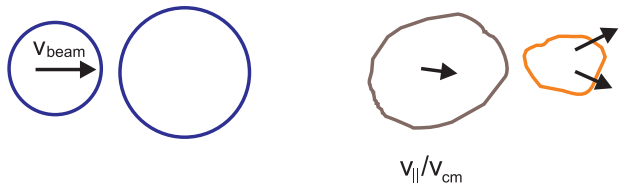}}
\caption{A mass-asymmetric collision.}
\label{asym}
\end{figure}

The stopping observables utilized for collisions include the
linear momentum
transfer (LMT) and ERAT.
In the LMT measurements,
central ($b \sim 0$) mass asymmetric reactions are assessed
within the laboratory frame, cf.\ Fig.\ \ref{asym}.
The velocity component along the beam of the most massive
fragment stemming from a reaction is identified, and its
average over
reaction events is compared to the cm velocity.  A proximity
of the average component to the net cm velocity, $\langle
v_\parallel \rangle \sim v_{cm}$, indicates fusion in a
reaction and, thus,
a large level of stopping and, potentially, large
elementary cross sections.  On the other hand, low values of
the average component, $\langle v_\parallel \rangle \sim 0$,
indicate little stopping and, potentially, low elementary cross
sections.

\begin{figure}
\centerline{\includegraphics[angle=90,
width=.65\linewidth]{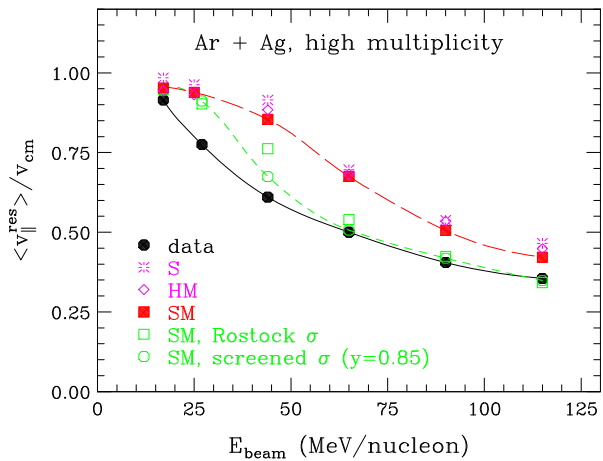}}
\caption{Measured (filled circles) and calculated (other
symbols) average velocity ratio $\langle v_\parallel
\rangle / v_{cm}$ as a function of beam energy in central
$^{40}$Ar + Ag collisions. }
\label{lmt}
\end{figure}

The Stony Brook group \cite{col98} has investigated central
($\langle
b \rangle \sim b_{max}/4$) collision events of Ar with several
targets, Cu, Ag and Au, and has determined $\langle v_\parallel
\rangle / v_{cm}$ as a function of bombarding energy; the
results
from the Ag target are represented by filled circles in
Fig.~\ref{lmt}.
At low energies, the
projectile and target appear to fuse.  As energy is raised,
the
transparency sets in and it increases with the increase in
energy.
Results of transport simulations assuming free nucleon-nucleon
cross sections and different forms of optical potentials
are represented, respectively, by stars, diamonds and filled
squares in the figure.  It is seen that all those calculations
overestimate
the stopping.  The fusion continues too high up in energy and
at high energies the residue velocity remains too high.
Notably, the results are rather insentive to the assumed form
of nucleon single-particle energies.  In consequence, these
results point to
the in-medium cross-sections reduced compared the free-space,
or increased viscosity.

There may be different reasons for an in-medium reduction of
cross sections.  Thus, it may be reasonable to assume that the
geometric cross-section radius should not exceed the
interparticle distance,
\beq
\sigma \lesssim y \, \rho^{-2/3} \, ,
\eeq
with $y \sim 1$, since, otherwise, the nucleon-nucleon
scatterings can get multiply counted.  The requirement may be
implemented in practice with the following in-medium cross
section:
\beq
\sigma = \sigma_0 \, \tanh{(\sigma_{free}/ \sigma_0)} \, ,
\hspace{2em} \mbox{where} \hspace{2em}
\sigma_0 = y \, \rho^{-2/3} \, .
\label{sigsc}
\eeq

There may be other reasons for the cross-section reduction,
such as the effects of Pauli principle and of single-particle
energy modifications for intermediate states.  In the
calculations that include those effects (but not the overlap of
binary collision regions), such as of the Rostock group
\cite{sch98}, a general reduction of
the in-medium cross sections is found.  In the following, we
utilize a crude parametrization of the Rostock cross sections:
\beq
\sigma = \sigma_{free} \, \exp{\left( - 0.6 \, {\rho \over
\rho_0} \, {1 \over 1 + (T_{cm}/150\, {\rm MeV})^2} \right)} \,
\label{sigros}
\eeq
where $T_{cm}$ is the c.m.\ kinetic energy of a scattering
nucleon pair.

The results of the simulations using the two types of reduced
in-medium cross-sections are shown in Fig.\ \ref{lmt} with open squares
and open circles, respectively.  It is seen that the stopping
is reduced now at higher energies and in a much better
agreement with data.

\begin{figure}
\centerline{\includegraphics[angle=90,
width=.67\linewidth]{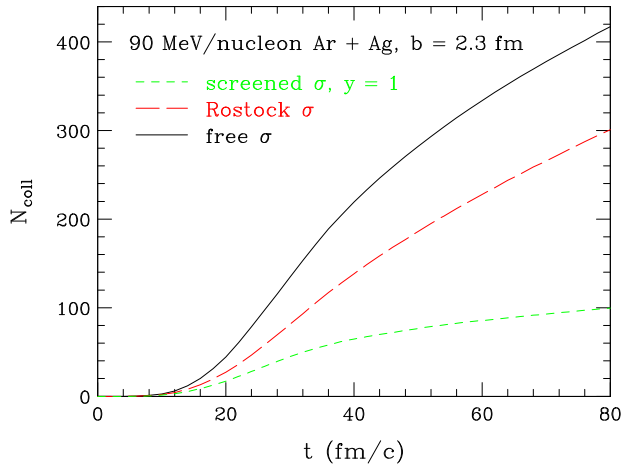}}
\caption{Number of collisions in the 90 MeV/nucleon Ar + Ag
reaction for different cross sections, as a function of time.}
\label{colno}
\end{figure}

While similar reductions are obtained with the two in-medium
cross sections, the two cross sections are actually quite
different.  This is illustrated in Fig.\ \ref{colno} that shows the
number of collisions for the different cross sections, as a
function of time.
It is seen that the number of collisions for the Rostock cross
sections is reduced by $\sim 25$\% compared to the free
cross-sections.  However, the number of collisions for the
cross sections screened with the interparticle distance is
reduced by a factor of~4.  How come those two cross sections
lead
to the same reduction in stopping when the collision numbers
are so vastly different?

Clearly, not all collisions are the same.  If e.g.\ the
scattering angle in collision is small,
the~collision may
matter little for the reaction dynamics.  Notably also such
collisions are
also most peripheral and most ambigous within a many-body
system.
In the expression for viscosity (\ref{eta}), the
collisions are
weighted with the weight
$q_{12}^4 \, \sin^2 \, \theta'$, suppressing the collisions at
low
scattering angle, and weighting most those that take place at
large relative momentum and lead to $\theta ' ~ 90^\circ$.

\begin{figure}
\centerline{\includegraphics[angle=90,
width=.67\linewidth]{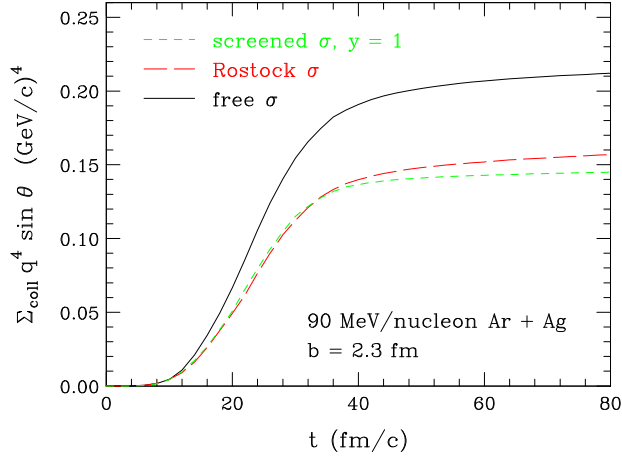}}
\caption{Number of collisions weighted with $q^4 \, \sin^2
{\theta}$ in the 90 MeV/nucleon Ar + Ag
reaction for different cross sections, as a function of time.}
\label{colnowg}
\end{figure}

While the two different parametrizations of cross sections
yield different results regarding the collision number, it is
interesting to ask whether they also yield different results
for collisions weighted with their importance in the
expression for viscosity.  This is examined in Fig.~\ref{colnowg} and it is
seen that the two parametrizations, that yield a right
reduction in stopping, also practically agree with regard to
the weighted collision number.  These parametrizations may be
then expected also to agree with regard to an (increased)
viscosity of the system.

\begin{figure}
\centerline{\includegraphics[angle=90,
width=.67\linewidth]{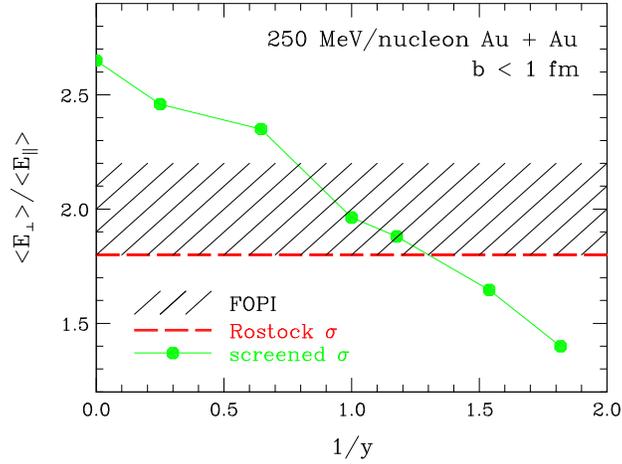}}
\caption{ERAT in central Au + Au reactions at 250 MeV/nucleon.
The filled circles represent the results of simulations as a
function of the parameter $1/y$ controlling the cross section
reduction in (\protect\ref{sigsc}).  The dashed line represents
the result of
simulations with Rostock cross sections (\protect\ref{sigros}).
The dashed region
represents the data of Ref.~\protect\cite{rei97}.}
\label{erat}
\end{figure}

Another nuclear stopping observable has
been the reaction
cross section for different values of $ERAT =
E_\perp/E_\parallel$,
examined in central Au + Au collisions by the FOPI
Collaboration \cite{rei97}.  Here, $E_\perp$
and $E_\parallel$ are the transverse and longitudinal energy,
respectively.  Generally, a
value of $ERAT < 2$ indicates a transparency (2~because of two
transverse dimensions and only one longitudinal), $ERAT > 2$
indicates a system splashing in the directions transverse to
the beam axis, and $ERAT=2$ indicates isotropy.  However,
finite-multiplicity fluctuations spread out and modify those
results and likewise do the detector inefficiencies.  After
correcting for the fluctuations and inefficiencies, the FOPI
Collaboration concluded that the head-on Au + Au collisions at
250 MeV/nucleon were consistent with isotropy.  Figure~\ref{erat} shows
the results for the expected value of $ERAT$ in simulations,
with the variation of the inverse of parameter~$y$ in the
first of our in-medium cross-section parametrizations, together
with the result for the second parametrization and for data
(with 10\% uncertainty).  The value of $1/y=0$ corresponds to
free cross sections and these again yield too much stopping.
The~compatibility with data requires $y \sim 1$.  In the
analysis, the Rostock and screened cross-section
parametrizations yield again very different collision numbers,
but similar numbers for collisions entered with viscous
weight, when those parametrizations yield a similar stopping.
The number of weighted collisions is again reduced by about
30\% compared to the case of free cross sections.

\begin{figure}
\centerline{\includegraphics[angle=0,
width=.76\linewidth]{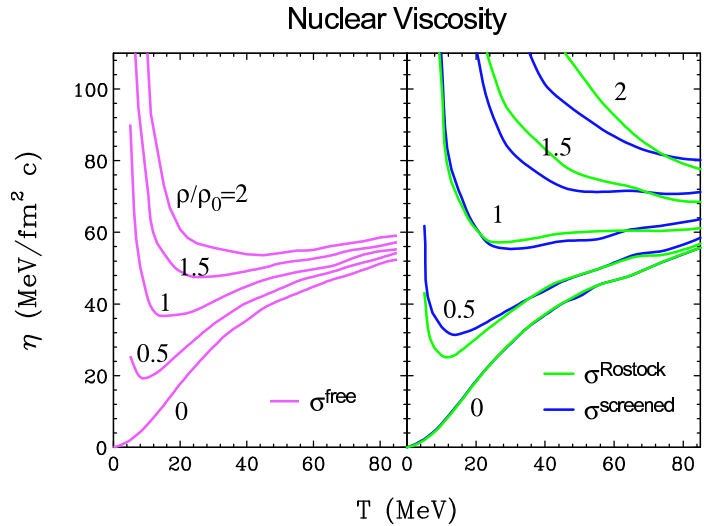}}
\caption{Viscosity in symmetric nuclear matter as a function of
temperature $T$ at different densities $\rho/\rho_0$ for free
NN cross sections (left panel) \protect\cite{dan84} and for
medium-modified cross sections (right panel).
}
\label{visc}
\end{figure}

Based on the simulations, we can conclude that the stopping
observables indicate reduced in-medium cross sections.  Details
of the reduction appear ambiguous but the stopping primarily
appears sensitive to the nuclear viscosity.
The very different parametrizations of the cross
sections that yield an agreement with data appear relatively
consistent with regard to the enhanced nuclear viscosity at
densities and temperatures such as explored in the reactions,
see Fig. \ref{visc}.
After tackling the viscosity and in-medium cross-sections,
we now turn to the features of the nuclear
EOS.

\section{Nuclear Incompressibility}

From the binding-energy formula and from electron scattering,
we
know that the energy per nucleon in symmetric nuclear matter,
under
the effects of nuclear forces alone, minimizes at the normal
density $\rho_0 = 0.16$~fm$^{-3}$ at -16~MeV.  The curvature
around the minimum is quantified in terms of
incompressibility~$K$, first introduced as a curvature of the
energy with respect to the nuclear radius for considered
sharp-sphere nuclei,
\beq
K = 9 \, \rho_0^2 \, \frac{d^2}{d \rho^2}
\left(\frac{E}{A}\right)
= R^2 \, \frac{d^2}{dR^2} \left(\frac{E}{A}\right) \, .
\eeq

\begin{figure}
\centerline{
\includegraphics[width=.52\linewidth]{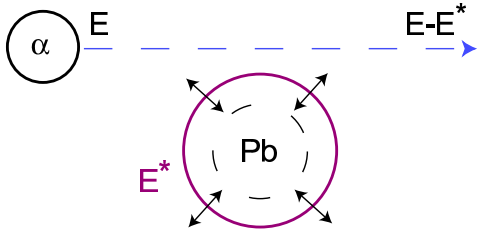}
}
\caption{Volume oscillations induced by alpha scattering.}
\label{alpha}
\end{figure}

The simplest way to determine the incompressibility
may seem to induce volume oscillations in a
nucleus.  This could be done by scattering $\alpha$ particles
off
a nucleus, Fig.~\ref{alpha}.  For the lowest excitation, the
excitation energy $E^*$, deduced from the final $\alpha$
energy, would be related to the classical frequency through
$E^*
= \hbar \Omega$, and the latter would be related to~$K$.
Let us examine the classical energy of an oscillating nucleus:
\bea
E_{tot} & = & \int d{\bf r} \, \rho \, \frac{m_N \, v^2}{2} +
\frac{1}{2} \, A \, K \, (R - R_0)^2\nonumber \\
& = & \frac{A m_N  \langle r^2 \rangle_A
\dot{R}^2}{2} + \frac{1}{2} \, A \, K \, (R - R_0)^2 \, ,
\eea
where we use the fact that, for a nucleus uniformly changing
its density, the velocity is proportional to the radius,
$v = \dot{R} \, (r/R)$.  We then obtain the energy of a simple
harmonic oscillator; the frequency is a square root of the
spring constant divided by mass constant, yielding:
\beq
E^* = \hbar \, \sqrt{\frac{K}{m_N \, \langle r^2 \rangle_A}}
\, .
\eeq

\begin{figure}
\centerline{ \includegraphics[width=.70\linewidth]{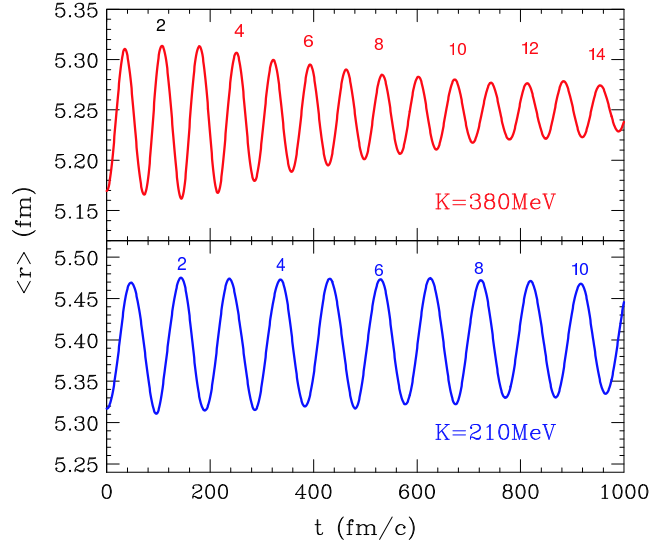} }
\caption{Radius of an expanded lead nucleus as a function of
time from the Vlasov version of (\ref{Boltzmann}), for two
values of incompressibility.}
\label{radius}
\end{figure}

\begin{figure}
\begin{center}
\parbox{.45\linewidth}
{
\includegraphics[angle=0,
width=1.06\linewidth]{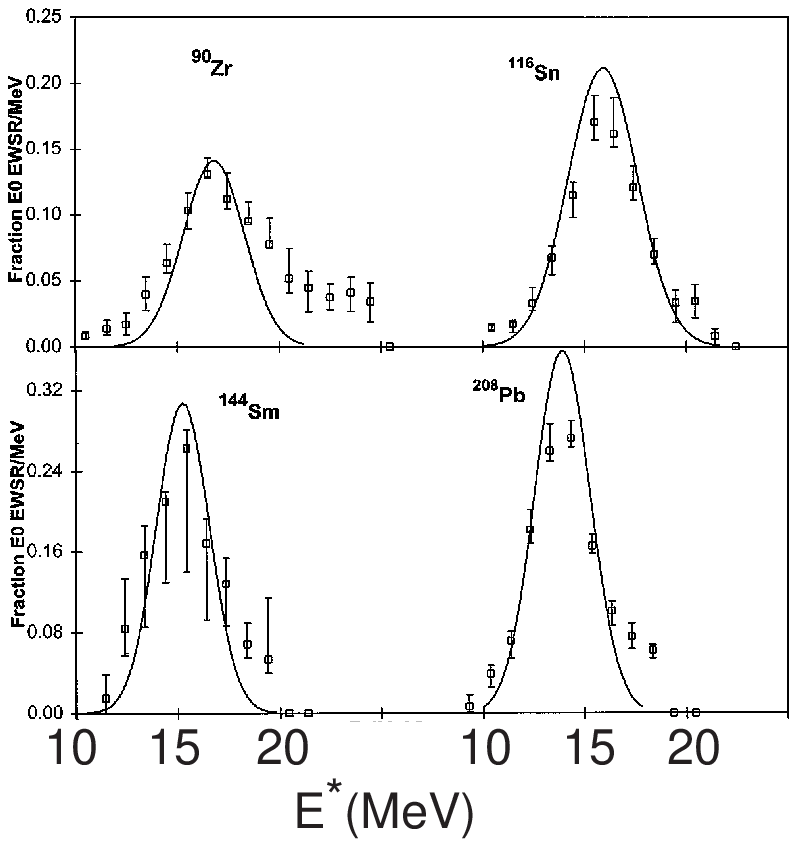}}
\parbox{.45\linewidth}
{\vspace*{.6in}
\hspace*{1.5em}\includegraphics[angle=0,
width=1.06\linewidth]{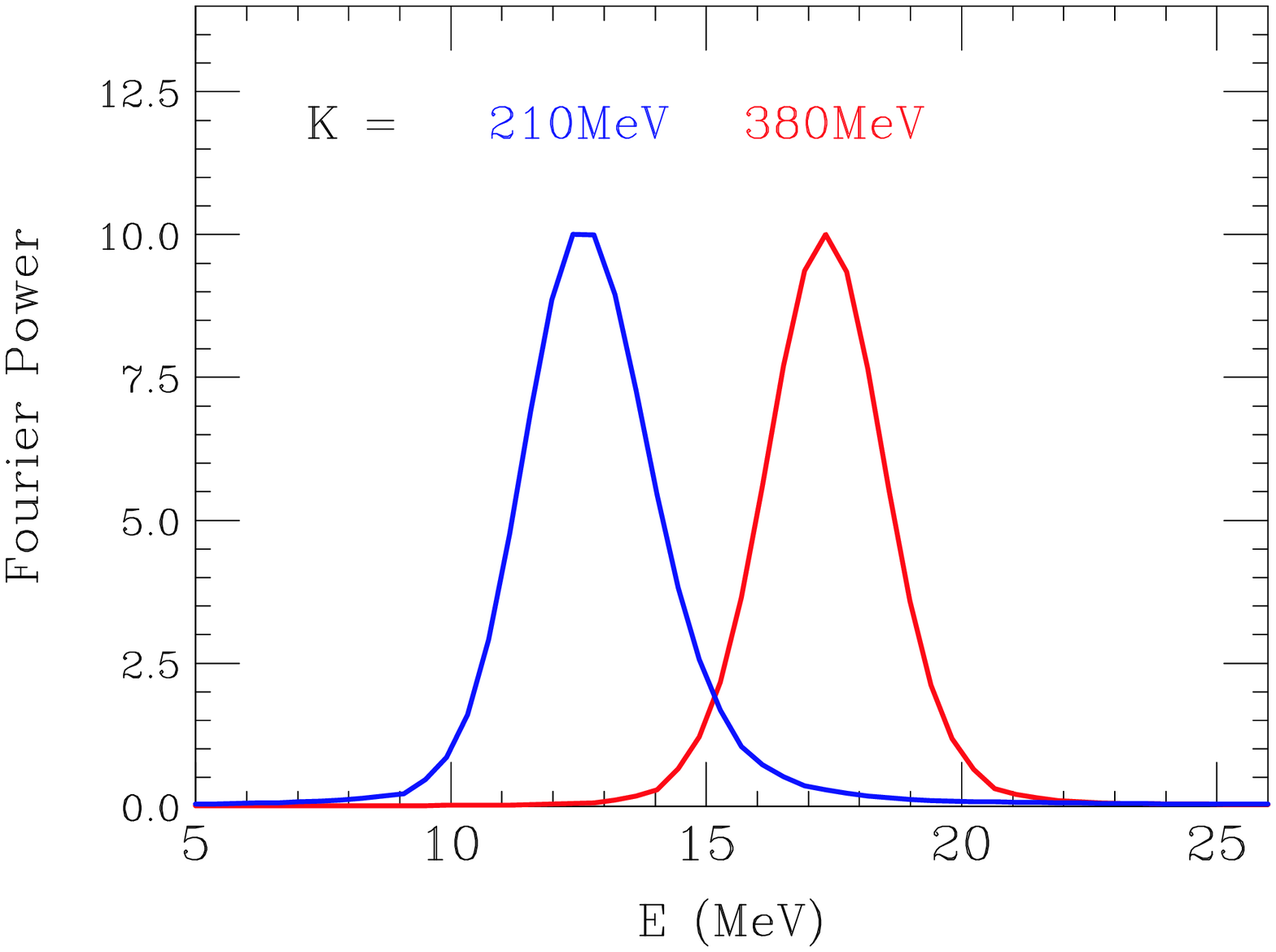}\\[-1ex]}
\end{center}
\caption{Left: $0^+$ excitation spectrum
in several nuclei from measurements of Youngblood
\protect\etal\ \protect\cite{you01}.
Right: Fourier spectrum for monopole oscillations in lead
within the Vlasov equation for two values of~$K$.}
\label{power}
\end{figure}

There are complications regarding this
reasoning.  Thus, the nucleus is not a sharp-edged sphere
and the Coulomb interactions play a role in the oscillations
in addition to the nuclear interactions different in isospin
asymmetric matter than in symmetric.  These effects
may be accounted for in time-dependent Hartree-Fock or in the
random-phase-approximation calculations allowing for meaningful
comparisons
to data.  The above approaches include also shell effects but,
if one wants to study just average features of excitations,
then the model based on (\ref{Boltzmann}) may be employed,
provided that
the net energy includes contributions from the finite-range of
interactions besides Coulomb, isospin and symmetric volume
terms \cite{dan00}.  If a nucleus is expanded, by increasing distances
from the center by a small fraction, then oscillations result,
illustrated in Fig.~\ref{radius}, with a distinct dependence
on~$K$.
Figure \ref{power} shows next the power spectrum for the oscillations
from the Boltzmann equation as well as the $0^+$ spectra from
precise analyses of alpha scattering \cite{you01}, in the scattering
angle and energy loss.
Next, Fig.~10 compares the mass dependence of the resonance
energy with the results from the Vlasov equation.  The data
favor $K = 225 \pm 15$~MeV, represented by the intermediate
line.

\begin{figure}
\begin{center}
\includegraphics[width=.67\linewidth,angle=0]{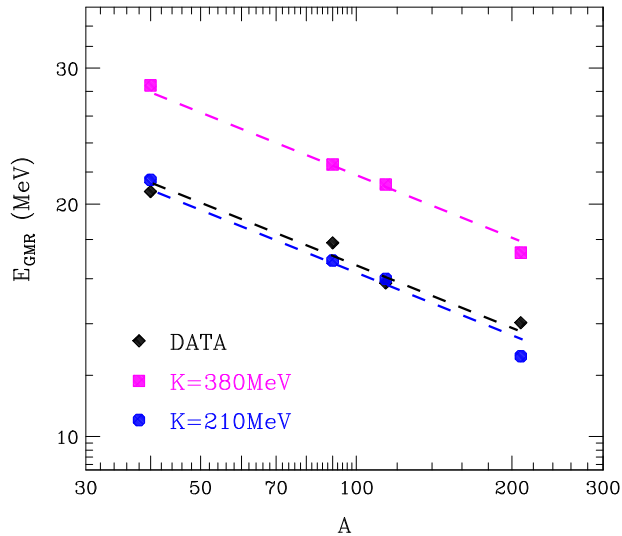}
\caption{Measured \protect\cite{you01} and calculated energies of
giant monopole resonances in spherical nuclei.}
\end{center}
\label{GMRA}
\end{figure}

\section{EOS at Supranormal Densities from Flow}

Features of EOS at supranormal densities can be inferred from
global features of flow in collisions of heavy nuclei at
high energies.  At low impact parameters, relatively
large regions of high density are formed and matter is
best equilibrated.
The collective flow can provide access to
pressure generated in the collision.

To see how the flow relates to pressure, we may look at the
hydrodynamic Euler equation for the nuclear fluid, an analog of
the Newton equation, in a local frame where the collective
velocity vanishes, $v=0$:
\beq
(e + p) \, \frac{\partial}{\partial t} \, \vec{v} =
- \vec{\nabla} p \, .
\eeq
The collective velocity becomes an observable at the end of the
reaction.  In comparing to the Newton equation, we see that the
pressure
$p= \rho^2
\frac{\partial(e/\rho)}{\partial \rho}|_{s/\rho}$ plays the
role
of a potential for the hydrodynamic motion, while the density
of enthalpy $w=e+p$ plays the role of a mass.  In fact, at
moderate energies, the enthalpy density is practically the mass
density, $w \approx \rho \, m_N$.  We see from the Euler
equation that the collective flow can tell us about the
pressure
in comparison to enthalpy.  In establishing the relation, we
need
to know the spatial size where the pressure gradients develop
and this will be determined by the nuclear size.  However, we
also
need the time during the hydrodynamic motion develops and this
can represent a problem.

Notably, the first observable that one may want to consider
to extract the information on EOS is the net radial or
transverse
collective energy. That energy may reach as much as half of the
total kinetic energy in a reaction.  Despite its magnitude, the
energy is not useful for extracting the information on EOS
because of the lack of information on how long
the energy develops.  Large pressures acting over a short
time
can produce the same net collective energy as low pressures
acting
over a long time.  This makes appearent the need for a timer in
reactions.

\begin{figure}
\centerline{\includegraphics[angle=0,
width=.92\linewidth]{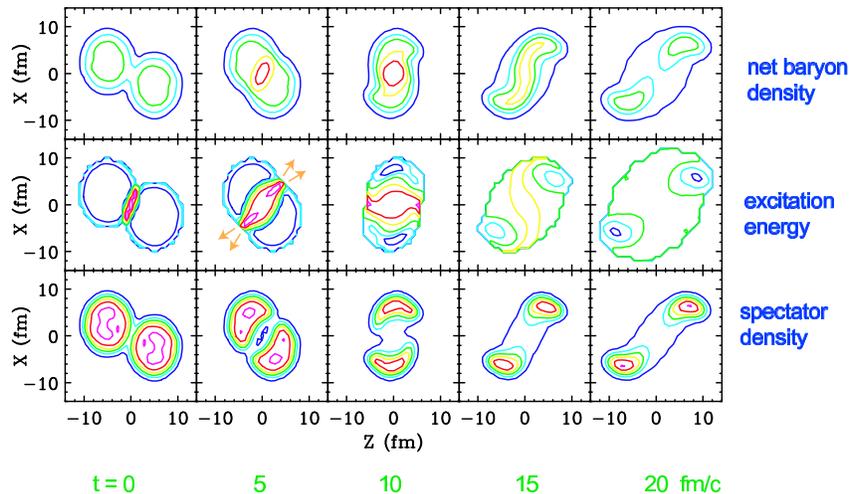}}
\caption{Reaction-plane contour plots for different quantities
in a $^{124}$Sn
+ $^{124}$Sn reaction at 800 MeV/nucleon and $b=6$~fm, from
transport simulations by Shi \protect\cite{shi01}.}
\label{contour}
\end{figure}

The role of the timer in reactions may be taken on by the
so-called spectators.  The spectator nucleons are those in the
periphery of an energetic reaction, weakly affected by the
reaction process, proceeding virtually at undisturbed original
velocity, see Fig.~\ref{contour}.  Participant
nucleons, on the other hand, are those closer to the center of
the reaction, participating in violent processes, subject to
matter compression and expansion in the reaction.  As the
participant zone expands, the spectators, moving at a
prescribed pace, shadow the expansion.  If the pressures in the
central region are high and the expansion is rapid, the
anisotropies
generated by the presence of spectators are going to be strong.
On the other hand, if the pressures are low and,
correspondingly, the expansion of the matter is
slow, the shadows left by spectators will not be very
pronounced.

There are different types of anisotropies in the emission that
the spectators can produce.  Thus, throughout the early stages
of
a collisions, the particles move primarily along the beam
axis in the center of mass.  However, during the compression
stage, the participants get locked within a channel, titled at
an angle, between the spectator pieces, cf.~Fig.~\ref{contour}.
As a consequence,
the forward and backward emitted particles acquire an
average deflection away from the beam axis, towards the
channel direction.  Another anisotropy is the ellipticity
$v_2$, that
we already examined as a function of $p^\perp$ in midperipheral
collisions.  Now we will consider global $v_2$ values at lower
impact parameters.

\begin{figure}
\centerline{\includegraphics[angle=0,
width=.72\linewidth]{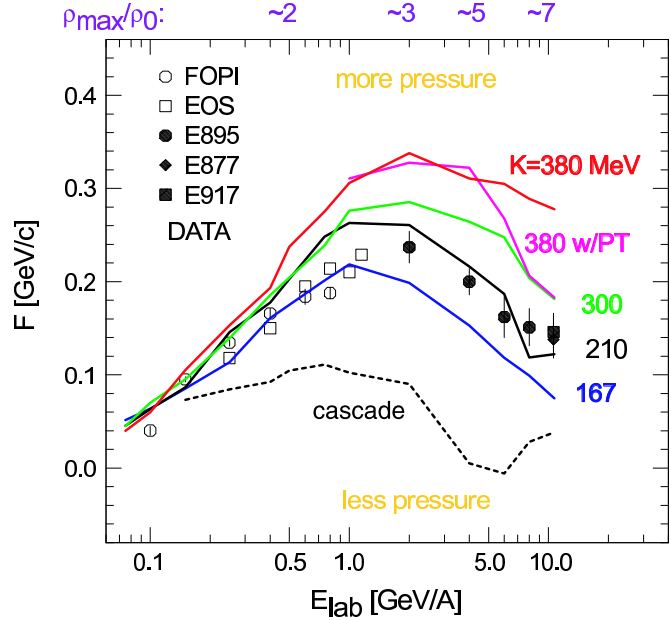}}
\caption{Sideward flow excitation function for Au + Au.  Data
and transport calculations are respresented, respectively, by
symbols and lines \protect\cite{dan01}.}
\label{flow}
\end{figure}

The different anisotropies have been quantified experimentally
over a wide range of bombarding energies in Au + Au collisions.
Figure~\ref{flow}
shows the measure of the sideward forward-backward deflection
as a function of the beam energy, with
symbols representing data.  Lines represent
simulations assuming different EOS.  On top of the figure,
typical maximal densities are indicated which are reached at a
given bombarding energy.
Without interaction
contributions to pressure, the simulations labelled
cascade produce far too weak anisotropies to be compatible with
data.  The simulations with EOS characterized by the
incompressibility $K=167$~MeV yield adequate anisotropy at
lower beam energies, but too low at higher
energies.  On the other hand, with the EOS characterized by
$K=380$~MeV, the anisotropy appears too high at virtually all
energies.  It should be mentioned that the incompressibilities
should be considered here as merely labels for the different
utilized EOS.  The
pressures resulting in the expansion are produced at densities
significantly higher than normal and, in fact, changing in the
course of the reaction.

\begin{figure}
\centerline{\includegraphics[angle=0,
width=.72\linewidth]{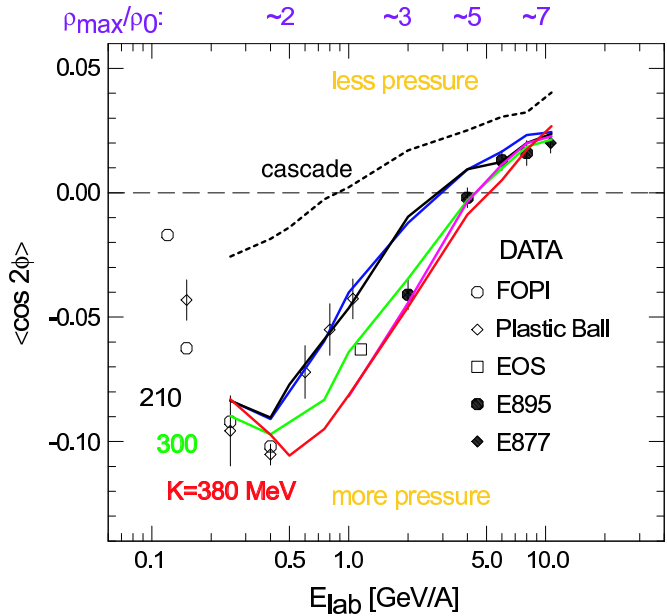}}
\caption{Elliptic flow excitation function for Au + Au.  Data
and transport calculations are respresented, respectively, by
symbols and lines \protect\cite{dan01}.}
\label{v2}
\end{figure}

Figure~\ref{v2} shows next the anisotropy of emission at
midrapidity, with symbols representing data and
lines representing simulations.  Again, we see that without
interaction
contributions to pressure, simulations cannot reproduce the
measurements.  The simulations with $K=167$~MeV give too little
pressure at high energies, and those with $K=380$~MeV generally
too much.  A level of discrepancy is seen between data from
different experiments.

\begin{figure}
\centerline{\includegraphics[angle=0,
width=.72\linewidth]{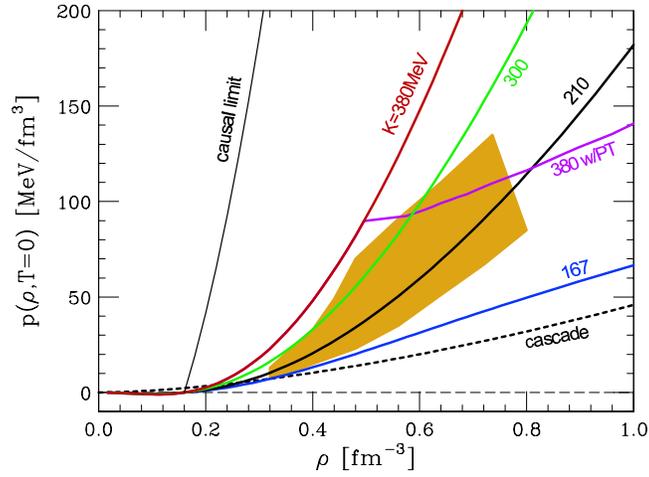}}
\caption{Constraints from flow on the $T=0$ pressure-density
relation, indicated
by the shaded region \protect\cite{dan01}.}
\label{Prho}
\end{figure}

\begin{figure}
\centerline{\includegraphics[angle=0,
width=.72\linewidth]{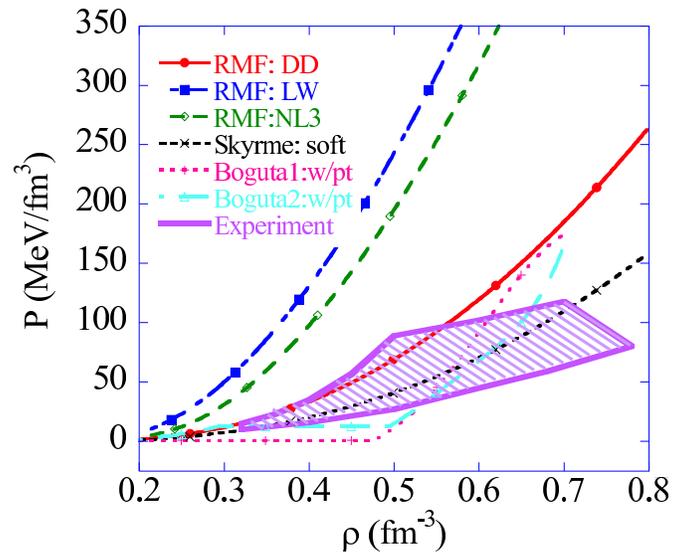}}
\caption{Impact of the constraints on models for EOS \protect\cite{dan01}.
}
\label{lynch}
\end{figure}

We see that no single EOS allows for a simultaneous description
of both types of anisotropies at all energies.  In particular,
the $K=210$~MeV EOS is the best for the sideward anisotropy,
and the $K=300$~MeV EOS is the best for the
elliptic anisotropy.  We can use the discrepancy between the
conclusions drawn from the two types of anisotropies as a
measure of inaccuaracy of the theory and draw broad boundaries
on pressure as a function of density from what is
common in conclusions based on the two anisotropies.
To ensure
that the effects of compression dominate in the reaction over
other effects,
we limit
ourselves to densities higher than twice the normal.  The
boundaries on the pressure are shown in
Fig.~\ref{Prho} and they eliminate some of the more extreme
models for EOS utilized in nuclear physics, such as the
relativistic NL3 model and models assuming a phase transition
at relatively low densities, cf.~Fig.~\ref{lynch}.

\section{Conclusions and Outlook}

Comparisons of transport model calculations to data can
yield information on bulk nuclear properties.
However, the progress has been difficult due to the need to
sort out competing physical effects.  Optimal observables are
those which are mostly sensitive to one uncertain nuclear
property.

Though the stopping observables are sensitive to the
in-medium cross sections, they probe cross sections weigthed
with scattering angle,
such as appear in the expression for nuclear viscosity.  These
appear reduced in lower-energy reactions by $\sim 30$~\%
compared to free space and the nuclear viscosity appears
increased respectively by $\sim 50$~\% compared to that
calculated with free cross sections.

Most straightforward determination incompressibility is by
analyzing
the excitation of density oscillations.  The far more
precise measurements of giant monopole resonances than in the
past suggest a value $K \sim 225$~MeV.

The flow in energetic reactions allows to place
meanigful constraints
on the nuclear pressure within the density range $2 \lesssim
\rho/\rho_0 \lesssim 5$.  The most extreme models for EOS can
be eliminated.

\section*{Acknowledgement}

This work was partially supported by the National Science
Foundation under Grant PHY-0070818.

\end{document}